\documentclass[twocolumn]{aastex631}

\newcommand{\plot}[1]{\textit{#1}:}
\newcommand{\code}[1]{\texttt{#1}}

\newcommand{\rh}{$r_{\text{h}}$}
\newcommand{\ebv}{$E(B-V)$}
\newcommand{\Rv}{$R_{V}$}
\newcommand{\Msun}{$\textup{M}_\odot$}
\newcommand{\Lsun}{$\textup{L}_\odot$}
\newcommand{\sqdeg}{deg$^2$}
\newcommand{\kms}{km\,s$^{-1}$}

\newcommand{\boov}{Bo\"otes V}

\def\ii{{~\sc ii}}
\def\ie{{i.e.,\ }}
\def\eg{{e.g.,\ }}

\usepackage[T1]{fontenc}
\usepackage{tipa}
\usepackage{accents}
\newcommand{\ubar}[1]{\underaccent{\bar}{#1}}

\shorttitle{Bo\"otes V}
\shortauthors{Smith et al.}
\graphicspath{{./}{figures/}}

\begin{document}

\title{Discovery of a new Local Group Dwarf Galaxy Candidate in UNIONS: Bo\"{o}tes V}

\author[0000-0002-6946-8280]{Simon E.T. Smith}
\affiliation{Department of Physics and Astronomy, University of Victoria, Victoria, BC, V8P 1A1, Canada}

\author{Jaclyn Jensen}
\affiliation{Department of Physics and Astronomy, University of Victoria, Victoria, BC, V8P 1A1, Canada}

\author{Joel Roediger}
\affiliation{NRC Herzberg Astronomy and Astrophysics, 5071 West Saanich Road, Victoria, BC, V9E 2E7, Canada}

\author{Federico Sestito}
\affiliation{Department of Physics and Astronomy, University of Victoria, Victoria, BC, V8P 1A1, Canada}

\author{Christian R. Hayes}
\affiliation{NRC Herzberg Astronomy and Astrophysics, 5071 West Saanich Road, Victoria, BC, V9E 2E7, Canada}

\author{Alan W. McConnachie}
\affiliation{NRC Herzberg Astronomy and Astrophysics, 5071 West Saanich Road, Victoria, BC, V9E 2E7, Canada}
\affiliation{Department of Physics and Astronomy, University of Victoria, Victoria, BC, V8P 1A1, Canada}

\author{Jean-Charles Cuillandre}
\affiliation{AIM, CEA, CNRS, Universit\'e Paris-Saclay, Universit\'e Paris, F-91191 Gif-sur-Yvette, France}

\author{Stephen Gwyn}
\affiliation{NRC Herzberg Astronomy and Astrophysics, 5071 West Saanich Road, Victoria, BC, V9E 2E7, Canada}

\author{Eugene Magnier}
\affiliation{Institute for Astronomy, University of Hawaii, 2680 Woodlawn Drive, Honolulu HI 96822}

\author{Ken Chambers}
\affiliation{Institute for Astronomy, University of Hawaii, 2680 Woodlawn Drive, Honolulu HI 96822}

\author{Francois Hammer}
\affiliation{GEPI, Observatoire de Paris, University\'e PSL, CNRS, Place Jules Janssen F-92195, Meudon, France}

\author{Mike Hudson}
\affiliation{Department of Physics and Astronomy, University of Waterloo, 200 University Ave W, Waterloo, ON N2L 3G1, Canada}
\affiliation{Waterloo Centre for Astrophysics, University of Waterloo, 200 University Ave W, Waterloo, ON N2L 3G1, Canada}
\affiliation{Perimeter Institute for Theoretical Physics, 31 Caroline St. North, Waterloo, ON N2L 2Y5, Canada}

\author{Nicolas Martin}
\affiliation{Universit\'e de Strasbourg, CNRS, Observatoire astronomique de Strasbourg, UMR 7550, F-67000 Strasbourg, France}
\affiliation{Max-Planck-Institut f\"ure Astronomie, K\"onigstuhl 17, D-69117, Heidelberg, Germany}

\author{Julio Navarro}
\affiliation{Department of Physics and Astronomy, University of Victoria, Victoria, BC, V8P 1A1, Canada}

\author{Douglas Scott}
\affiliation{Department of Physics \& Astronomy, University of British Columbia, Vancouver, BC, V6T 1Z1, Canada}



\begin{abstract}
    We present the discovery of \boov, a new ultra-faint dwarf galaxy candidate. This satellite is detected as a resolved overdensity of stars during an ongoing search for new Local Group dwarf galaxy candidates in the UNIONS photometric dataset. 
    It has a physical half-light radius of 26.9$^{+7.5}_{-5.4}$\,pc, a $V$-band magnitude of $-$4.5 $\pm$ 0.4\,mag, and resides at a heliocentric distance of approximately 100\,kpc. We use Gaia DR3 astrometry to identify member stars, characterize the systemic proper motion, and confirm the reality of this faint stellar system. 
    The brightest star in this system was followed up using Gemini GMOS-N long-slit spectroscopy and is measured to have a metallicity of [Fe/H] $=$ $-$2.85 $\pm$ 0.10\,dex and a heliocentric radial velocity of $v_r$ = 5.1 $\pm$ 13.4\,\kms. \boov\ is larger (in terms of scale radius), more distant, and more metal-poor than the vast majority of globular clusters. It is likely that \boov\ is an ultra-faint dwarf galaxy, though future spectroscopic studies will be necessary to definitively classify this object.
\end{abstract}



\section{Introduction} 
\label{sec:intro}
The discovery and study of dwarf galaxies in the Local Group has prospered due to the great advances in wide field imaging surveys over the last two decades \citep[e.g.][]{Willman2005a, Willman2005b, Belokurov2006, Belokurov2007, Belokurov2008, Laevens2015a, Laevens2015b, Koposov2015a, Torrealba2016, Torrealba2018, Torrealba2019, Mau2020, Cerny2021a, Cerny2021b, Cerny2022a, Sand2022}. The Sloan Digital Sky Survey \citep[SDSS;][]{York2000, Abazajian2009} led the early charge, with the Pan-STARRS consortium \citep{Chambers2016}, the Dark Energy Survey \citep[DES;][]{Abbot2018}, and the DECam Local Volume Exploration Survey \citep[DELVE;][]{Drlica2021} contributing mightily to the discovery of Milky Way satellites, while the Pan-Andromeda Archaeological Survey \citep[PAndAS;][]{McConnachie2009} has been prolific in finding dwarf galaxies around M31 \citep[e.g. ][]{McConnachie2008, Richardson2011, Martin2013}. The dwarf galaxy population around M31 has also been bolstered by more recent work with the DESI Legacy Imaging Survey \citep{Dey2019, Collins2022, Martinez2022}.

Dwarf galaxies play several key roles in testing and developing new models for answering some of the most fundamental questions in astronomy. Dwarf galaxies fainter than around $M_V = -7.7$ have been dubbed ``ultra-faint dwarf galaxies'' \citep[UFDs; e.g., see review by][]{Simon2019}. They appear to be the smallest, least massive, and most metal-poor galaxies yet observed, and so represent the extreme end of the galaxy luminosity function. They appear to reside in the shallowest gravitational potential wells that have been able to retain gas and stars throughout cosmic time \citep[e.g. ][]{Bovill2009, Simon2019}, and their existence tests current theories of galaxy formation, particularly the interplay between stellar feedback and gas retention \citep{Bullock2017}, the evolution of faint galaxies in the Milky Way environment \citep[e.g.][]{Wheeler2015}, and the quenching effects of reionization \citep[e.g. ][]{Bullock2000}. Critically, UFDs are powerful probes of theories of structure evolution, as they provide parsec-scale tests of dark matter models that were initially developed to explain the largest scale observations of the Universe \citep{Lovell2012, Bullock2017}. Discoveries of new Local Group dwarf galaxies continue to provide both unique individual examples and an ever-growing statistical sample of faint systems that are testing these aforementioned fundamental theories of structure and galactic evolution.

In this paper, we detail the discovery and characterization of a new ultra-faint dwarf galaxy candidate, which we call \boov. In Section \ref{sec:methods} we summarise the dataset in which it was detected and the search algorithm that was used. In Section \ref{sec:props}, we characterize the structural parameters of the dwarf galaxy candidate, as well as its distance and luminosity. In Section \ref{sec:dynmet}, we identify member stars observed with Gaia, measure the proper motion of the system, and present first estimates of its metallicity and dynamics. Finally, in Section~\ref{sec:conc}, we discuss the classification of \boov\, and summarise our results. Additionally, we draw the attention of the reader to the work of \citet{Cerny2022b}, which presents an independent discovery of this new satellite in the DELVE dataset \citep{Drlica2021, Drlica2022}.

\section{Data and Detection}
\label{sec:methods}

\begin{figure*}
    \centering
    \includegraphics[width=0.8\linewidth]{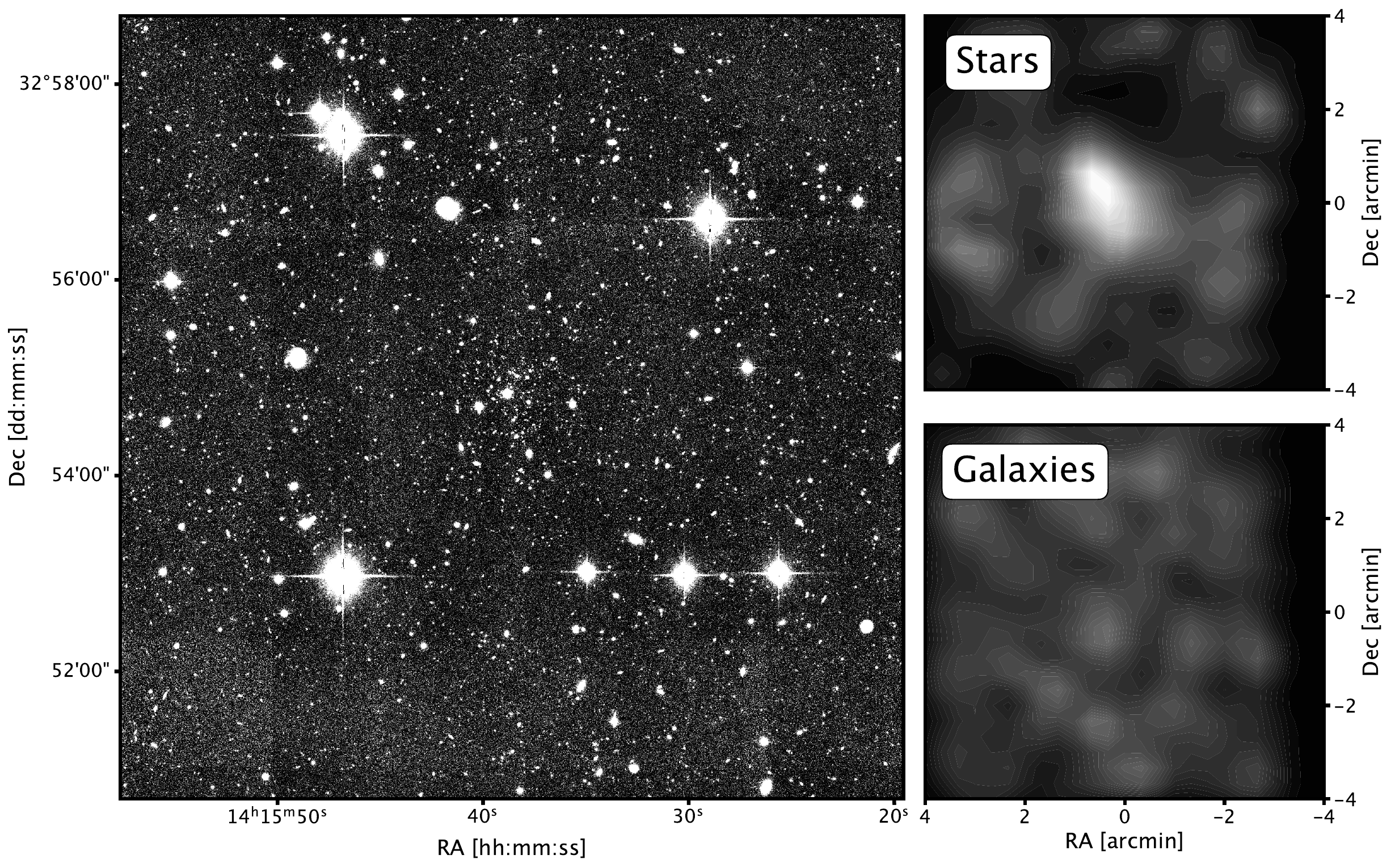}
    \caption{\plot{Left} 8\arcmin\,$\times$\,8\arcmin\ cutout from CFIS-r imaging tiles. Bright sources with diffraction spikes are likely Milky Way foreground stars. The clustering of stars and light in the centre of this image is \boov. \plot{Upper Right} Smoothed spatial density plot of all stellar sources on the 8\arcmin\,$\times$\,8\arcmin\ tile. No stellar population filtering has been applied to the stars in this smoothed distribution. \plot{Lower Right} Smoothed spatial density plot of all galaxies in the 8\arcmin\,$\times$\,8\arcmin\ tile, normalized with respect to the stellar density. There are many more galaxies than stars in this field of view, but there is no discernible overdensity.}
    \label{fig:tile+den}%
\end{figure*}

\subsection{UNIONS}
\label{subsec:unions}
\boov\ was identified as an overdensity of resolved stars in the Ultraviolet Near-Infrared Optical Northern Survey (UNIONS). UNIONS is a consortium of northern wide field imaging surveys, and consists of the Canada-France Imaging Survey (CFIS) collaboration that uses the Canada-France-Hawaii Telescope (CFHT), team members from Pan-STARRS, and the Wide Imaging with Subaru HyperSuprime-Cam of the Euclid Sky (WISHES). Each group is currently collecting imaging at their respective telescopes: CFHT/CFIS is targeting deep $u$ and $r$ band photometry, Pan-STARRS is obtaining deep $i$ and moderate/deep $z$ bands, and Subaru/WISHES is acquiring deep $z$ band. These independent efforts are directed, in part, to securing optical imaging to complement the Euclid space mission, although UNIONS is a separate consortium aimed at maximizing the science return of these large and deep ground-based surveys of the northern skies. When completed, the combined $ugriz$ survey will cover approximately 5000 \sqdeg\ at declinations of $\delta$ $>$ 30\textdegree\ and Galactic latitudes of $|b|$ $>$ 30\textdegree\ (the northern sky, excluding the Milky Way disk) and will be approximately as deep as one year of the Legacy Survey of Space and Time (LSST) at the Vera C. Rubin Observatory.

Our work uses the CFIS-r and Pan-STARRS-i combined dataset. For CFIS-r, the 5-sigma point source depth is 24.9 mag with a 2\,arcsecond (\arcsec) aperture. For Pan-STARRS-i, the final 5-sigma point source depth will be 24.3 mag. However, the Pan-STARRS survey strategy of mapping the entire sky repeatedly means that the i-band is currently, on average, at about 75\% of the final depth, although there is significant variation across the survey region. The area in common between both bands at this stage amounts to $\sim$3500\,\sqdeg\ total across both the North and South Galactic Caps (NGC/SGC). These catalogs were crossed matched using a 0.5\arcsec\ matching tolerance (although typically the sources match to better than 0.13\arcsec).

Star-galaxy separation was performed using morphological criteria in CFIS-r, which has a median image quality of better than 0.7\arcsec. We correct for the Galactic foreground extinction using the extinction values, \ebv, from \citet{Schlegel1998} assuming the conversion factors given by \citet{Schlafly2011} for a reddening parameter of \Rv\ = 3.1. For the CFIS-r, we adopt the conversion factor for the DES $r$-band: the full-width at half-maximum (FHWM) of the DES-r \citep{Flaugher2015} and CFIS-r are identical, and the DES $r$-band is shifted redwards with respect to the CFIS filter by only 2\,nm. 

\subsection{GMOS-N Spectroscopy}
\label{subsec:gmos}

We received Directors Discretionary Time at Gemini North to use the Gemini Multi-Object Spectrograph (GMOS-N) to obtain the spectrum of a single star, the brightest in \boov, through the program GN-2022B-DD-201 (PI: S. Smith). Our long-slit spectroscopic observations used the R831 grating and the RG610 spectroscopic blocking filter, with a 1.0\,\arcsec\ wide slit. We had a 1\,$\times$\,2 CCD binning configuration, resulting in a $\sim$\,0.38\,\AA\ per-pixel spectral resolution. Our observations were comprised of 3\,$\times$\,900\,s exposures centered at 8500\,\AA\ and 3\,$\times$\,900\,s exposures centered at 8600\,\AA\ during the nights of August 7th and August 8th, bringing our total exposure time to 5400\,s. 

The science observations were reduced and extracted using standard routines and procedures for GMOS data in the Gemini \textsc{IRAF} package. Bias corrections and flat-fielding were performed, along with the calibration of the wavelength solution using CuAr emission lamp exposures that were taken alongside the science observations. The three frames obtained at each central wavelength were stacked using \textsc{GemCombine}, and a one-dimensional spectrum was extracted from each stack. The flux was then calibrated using a 60\,s exposure of a spectrophotometric standard (G191B2B) using an identical instrument set-up with the observations centered at a wavelength of 8500\,\AA. The 8500\,\AA-centered stack and 8600\,\AA-centered stack were observed on consecutive nights, so their extracted spectra were individually corrected for Earth's orbital motion about the Sun before they were added together, producing the final spectrum of the target star. 

The exposure times, instrument set-up, and wavelength range were selected with the goal of measuring a signal-to-noise ratio (S/N) of $\sim$\,50 per-pixel in the region of the Calcium\ii{} infrared triplet (CaT) absorption feature. This has been demonstrated as sufficient to measure the heliocentric radial velocity with an uncertainty of $\sim$10 \kms\ \citep{Piatti2018}, and also sufficient to measure the CaT Equivalent Width (EW) with adequate precision for reaching an uncertainty of less than 0.2 dex in the metallicity estimate through the CaT EW - [Fe/H] relation of \citet{Starkenburg2010}. Details of the measured properties of this spectrum are found in Section \ref{subsec:spec}.

\subsection{Detection}
\label{subsec:detection}

\begin{figure*}
    \centering
    \includegraphics[width=0.8\linewidth]{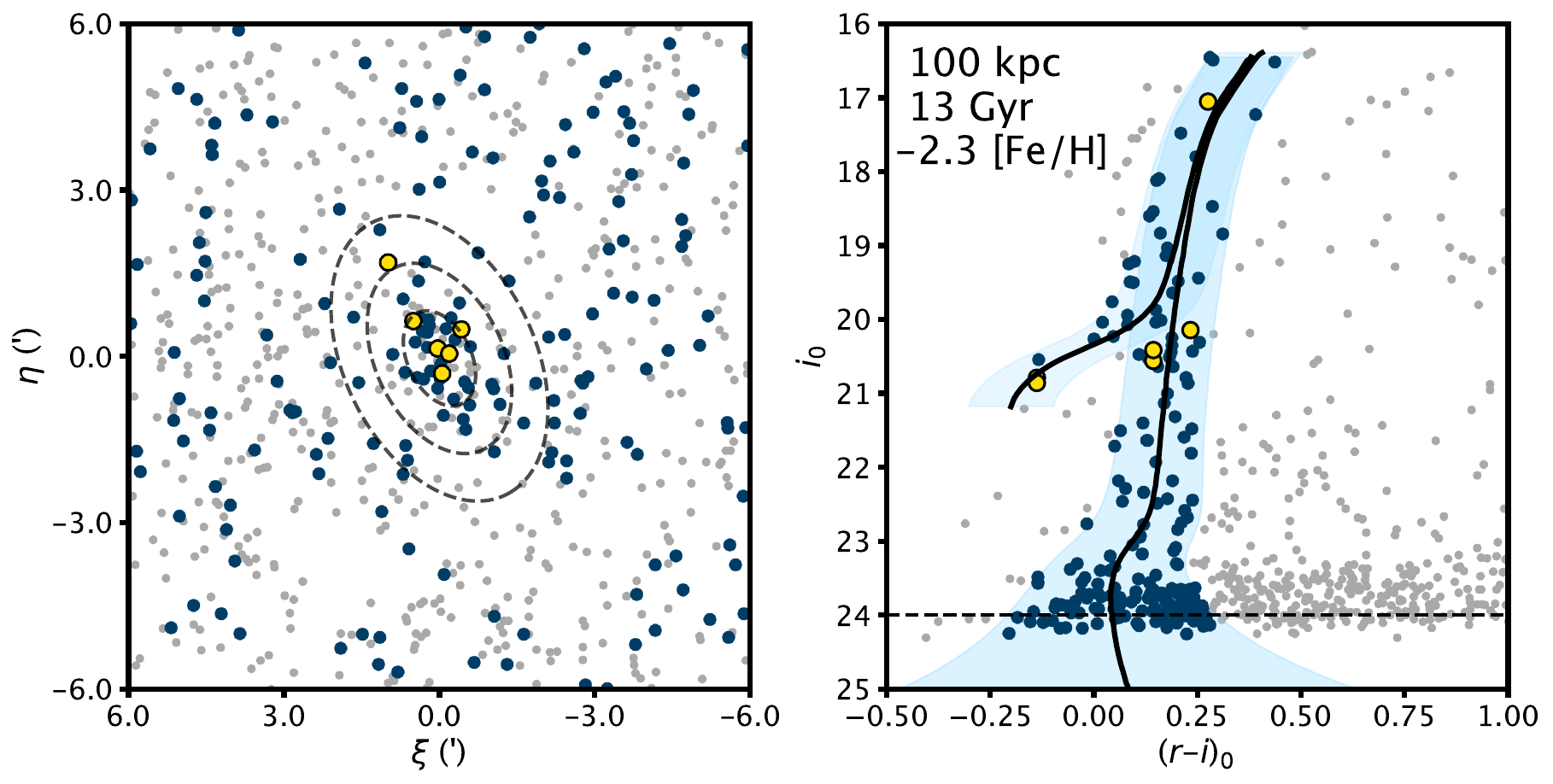}
    \caption{
    \plot{Left} Sky positions of all stars in a 12\arcmin\,$\times$ 12\,\arcmin\ region about \boov, projected onto the tangent plane centered at RA, Dec = (14h 15m 38.6s, +32\textdegree\ 54\arcmin\ 42\arcsec). Dark blue sources are those that meet the isochrone selection criterion (Right). Yellow points are stars identified in Gaia DR3 as sources with complete astrometric information, and are selected as high confidence (probability $>$ 90\%) members by our maximum-likelihood membership selection algorithm (see Section \ref{subsec:gaia}). The concentric black ellipses indicate 1\,$\times$, 3\,$\times$, and 5\,$\times$ the half-light radius (\rh) as determined by the MCMC fit (see Section \ref{subsec:structural}).
    \plot{Right} Color-magnitude diagram of extinction-corrected CFIS-r and Pan-STARRS-i for all stellar sources in a 12\arcmin\,$\times$\,12\arcmin\ region about \boov. We overlay an old (13\,Gyr), metal-poor ([Fe/H] = $-$2.3) isochrone, shifted to a distance of 100\,kpc. We use a broad color selection criterion, selecting all stars with $(r-i)_0 \leq 0.1$ from the isochrone and we add, in quadrature, the empirical photometric errors from each band. Stars consistent with this selection criterion are highlighted in dark blue. 
    }
    \label{fig:cmdselect}
\end{figure*}

\boov\  was identified as one of the most promising candidates in a new and ongoing search for dwarf galaxies in the UNIONS sky. The method of detection is based on a matched-filter approach, a tried and tested methodology that has yielded many previous dwarf galaxy discoveries \citep{Martin2008, Bechtol2015, Drlica2015}. We start by selecting all stars that are consistent with a 12\,Gyr, [Fe/H] = $-$2\,dex isochrone shifted to a heliocentric distance $d$. The algorithm filters the sky systematically in logarithmic steps for $d$ in the range [10, 1000] kpc and the isochrones used in this work were obtained from the PARSEC isochrone database \citep{Bressan2012} for the CFHT and Pan-STARRS 1 (PS1) photometric systems. We adopt a broad color cut around the isochrone that is more than sufficient to account for the empirical photometric errors in both the CFIS-r and Pan-STARRS-i, the intrinsic color spread of any putative dwarf galaxy, and the unknown distance of the dwarf. Formally:
    
\begin{equation}
    \big|(r-i)_{\text{star}} - (r-i)_{\text{iso}}\big| \leq \sqrt{0.1^2 + \sigma_r^2 + \sigma_i^2},
\end{equation}

\noindent where $(r-i)_{\text{star}}$ is the color of a given star, $(r-i)_{\text{iso}}$ is the color of the isochrone at the same magnitude as the given star, and $\sigma_{r,i}$ are the photometric uncertainties in each passband at the same magnitude as the given star. The algorithm does not consider stars fainter than CFIS-r = 24.6 and Pan-STARRS-i = 24, since below these limits completeness becomes an issue, although all detected stars are used when visually inspecting candidates. Stars meeting this color criteria are selected and projected onto the tangent plane, where they are spatially binned into 0.5\arcmin\,$\times$\,0.5\arcmin\ pixels. This pixel size was selected to be slightly smaller than the typical angular size of the smallest known dwarfs so that a single galaxy will appear as several pixels in the two dimensional (2D) stellar density map.

The next step concerns teasing out local overdensities of stars. We find the mean local stellar density for each pixel by convolving the field with a 2D top hat filter with a width of 20\arcmin, where the large kernel was chosen to sufficiently smooth the stellar density around a putative dwarf galaxy. We find the standard deviation in the local stellar density by taking the square root of the variance of these convolved maps. By definition, this is the square of the mean local density minus the mean of the squared local density, i.e.,

\begin{equation}
    \sigma_{\text{loc}} = \sqrt{\langle \rho_{\text{loc}} \rangle^2 - \langle \rho_{\text{loc}}^2 \rangle}~.
\end{equation}

\noindent $\rho_{\text{loc}}$ is the density of the local field and $\sigma_{\text{loc}}$ is the standard deviation in the local field. 

Returning to the unsmoothed distributions, we convolve these maps with 2D Gaussian kernels of different sizes, corresponding to dwarf galaxies with different projected sizes. Specifically, we adopt kernels with a FWHM of 1.2, 2.4, and 4.8\arcmin. We refer to these ``smoothed densities'' as $\rho_{\text{sm}}$. Finally, for the entire UNIONS sky, we compute a 2D map of the statistical overdensities with respect to the local stellar density as 

\begin{equation}
    s = \frac{\rho_{\text{sm}} - \rho_{\text{loc}}}{\sigma_{\text{loc}}},
\end{equation}

\noindent where $s$ is the significance of the overdensity in units of $\sigma_{\text{loc}}$. This process is repeated for a range of distances, where each distance produces three maps, one for each of the smoothing kernels.

During initial tests of this method, nearly all known dwarf galaxies in the UNIONS footprint have been detected with high significance. As a reference point, Canes Venatici II is detected with high significance at several distances. At a distance of 100\,kpc, with the stellar density map smoothed by a 1.2\arcsec\ kernel, Canes Venatici II (true distance of 160\,kpc) has $s = 2.8$, while \boov\ has a significance value of $s = 4.9$. \boov\ is detected strongly at many distances less than 150\,kpc and is one of the most prominent overdensities of match-filtered stars that has not previously been identified as a real stellar system. However, we note the work of \citet{DarraghFord2021}, who have developed a wavelet-based algorithm to search for dwarf galaxy candidates in the Gaia dataset. \boov\ was identified as a ``gold standard'' candidate in position-proper motion space, though their work did not present a follow-up study.

In support of the match-filter detection, we present the CFIS-r imaging of the putative dwarf galaxy (left panel of Figure \ref{fig:tile+den}) wherein a central clustering of sources is seen prominently. As mentioned in Section \ref{subsec:unions}, we separate sources in the imaging into stars and galaxies, and in the right panels of Figure \ref{fig:tile+den}, we show the surface density of both of these. Clearly, the clustering is not due to background galaxies.

We show the spatial density of the match-filter-selected stars alongside the color-magnitude diagram (CMD) of sources in the vicinity of \boov\ in Figure \ref{fig:cmdselect}. The positions of all these sources on the CMD suggest that many of the stars in this field of view are consistent with the same old, metal-poor stellar population given a putative distance to the system of 100\,kpc.

\section{Photometric and Structural Properties}
\label{sec:props}
\begin{figure}
    \centering
    \includegraphics[width=\linewidth]{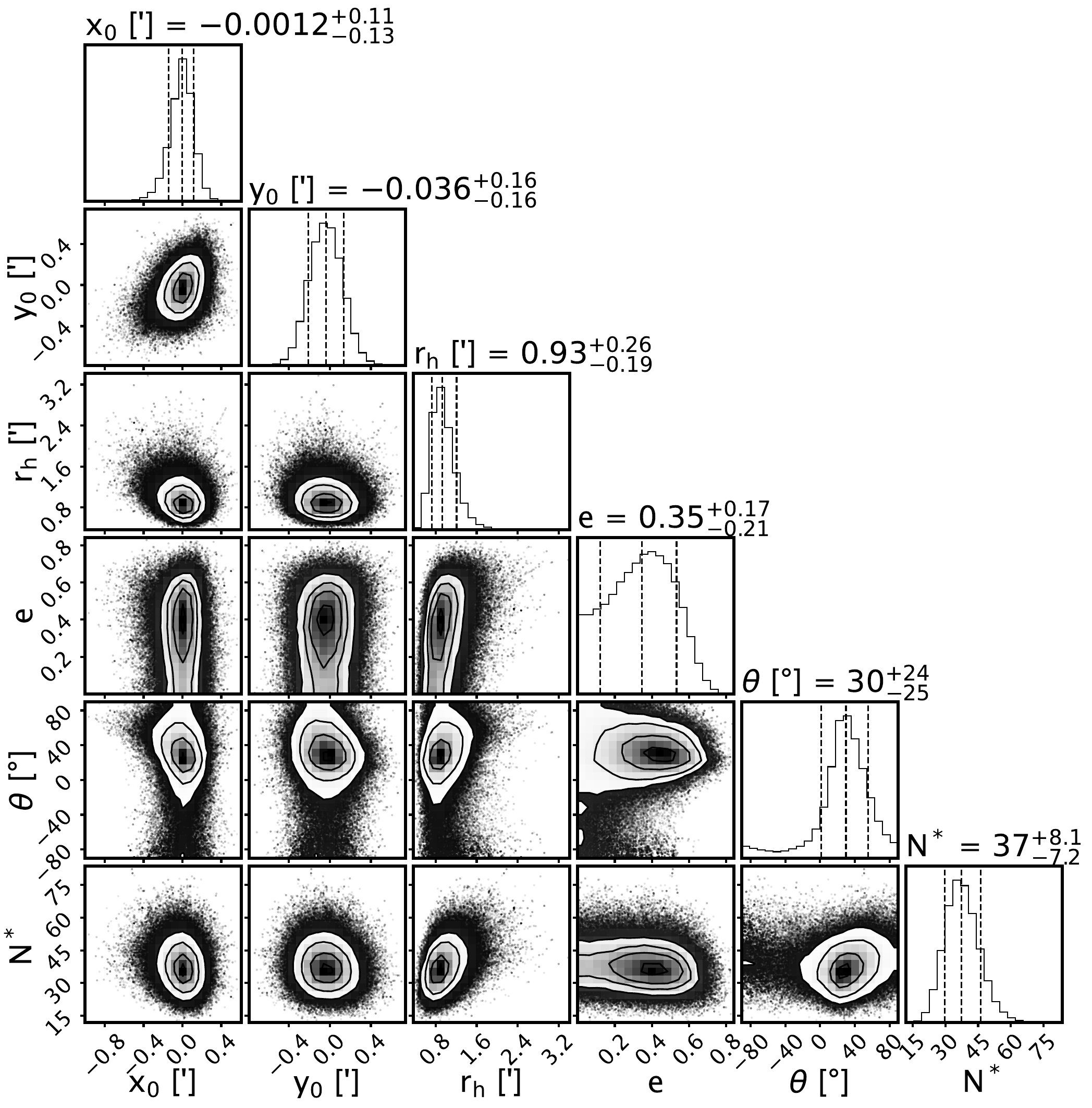}
    \caption{2D and marginalized PDFs from the MCMC analysis. The fit was run on all stellar sources within a 9\arcmin\-radius, circular region around the initial satellite's centroid estimate that are consistent with the isochrone selection criteria. This corresponds to all the stars which are highlighted in blue in Figure \ref{fig:cmdselect}. The quantities $x_0$, $y_0$, and \rh\ are the projected RA centroid offset, projected Dec centroid offset, and half-light radius, respectively.
    All three are presented in arcminutes. Ellipticity is denoted as $\epsilon = 1 - b/a$, where $b/a$ is the minor-to-major axis ratio, while $\theta$ is the position angle (denoted East from North in degrees), and $N^*$ is the estimated number of member stars. In each marginalized PDF, dashed lines should be interpreted as the 16th, 50th, and 84th quantiles, from left to right.}
    \label{fig:corner}
\end{figure}

\begin{figure}
    \centering
    \includegraphics[width=\linewidth]{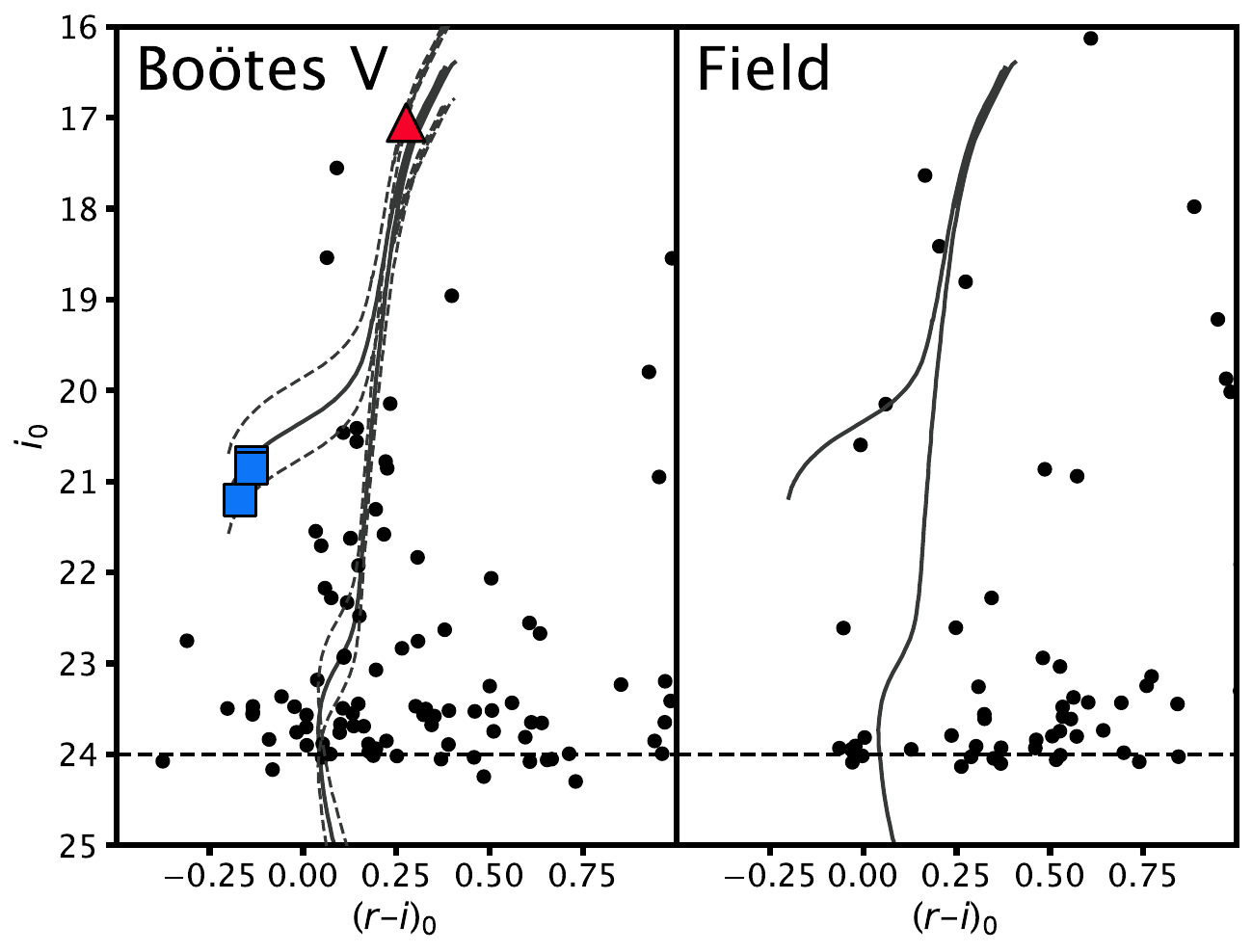}
    \caption{
    \plot{Left} CMD of all stars within 3 elliptical \rh\ of the satellite's centroid; 92 stars meet this spatial criterion. A 13\,Gyr, [Fe/H] = $-$2.3 isochrone shifted to 100\,kpc is overlaid as a solid line. Isochrones shifted to 80 and 120\,kpc, the lower and upper distance bounds as determined in Section \ref{subsec:dist}, are shown as dashed lines. A possible TRGB star (red triangle) and three possible BHB stars (blue squares), which were used for distance estimates, are shown as well. \textit{Note} - The two brightest BHB stars are nearly directly atop eachother on the CMD.
    \plot{Right} CMD of stars within an equivalent area offset by 12\arcmin\ to the south of the satellite's centroid, with the same isochrone overlaid; 48 stars meet this spatial criteria.}
    \label{fig:field}
\end{figure}

\subsection{Structure}
\label{subsec:structural}

Given the probable identification of a new Milky Way satellite as shown in Figure \ref{fig:cmdselect}, we determine the most likely values of its structural parameters using a Markov-Chain Monte Carlo (MCMC) approach under the assumption that the system is well described by an exponential model. 

Our model and approach are based on the work in \citet{Martin2008} and \citet{Martin2016}, and we used \code{emcee} \citep{ForemanMackey2013} to sample the posterior. The radial surface density profile $\rho_{\text{dwarf}}(r)$ for a dwarf galaxy can be described with an elliptical, exponential model as a function of $r$. This profile is defined by the centroid of the profile ($x_0, y_0$), an ellipticity $\epsilon$ (defined as $\epsilon = 1 - b/a$ where $b/a$ is the minor-to-major-axis ratio of the model), the position angle of the major axis $\theta$, (defined East of North), the half-light radius (which is the length of the semi-major axis \rh), and the number of stars $N^*$ in the system. The model is written as 

\begin{equation}
    \rho_{\text{dwarf}}(r) = \frac{1.68^2}{2\pi r_h^2 (1-\epsilon)}N^* \exp{\bigg(\frac{-1.68r}{r_{\text{h}}}\bigg)},
\end{equation}

\noindent where $r$, the elliptical radius, is related to the projected sky coordinates ($x$, $y$) by

\begin{eqnarray}
    \nonumber r = \Bigg\{ \bigg[\frac{1}{1-\epsilon} \Big((x-x_0)\cos\theta - (y-y_0)\sin\theta\Big) \bigg]^2 \\ + \bigg[ \Big((x-x_0)\sin\theta - (y-y_0)\cos\theta\Big)^2 \bigg]^2\Bigg\}^{\frac{1}{2}}.
\end{eqnarray}

We assume that the background stellar density is constant, which is reasonable on the scale of arcminutes up to a degree or so.  As such, we model the density of stars in our entire field of view as

\begin{equation}
    \rho_{\text{model}}(r) = \rho_{\text{dwarf}}(r) + \Sigma_{\text{b}},
\end{equation}

\noindent where $\Sigma_{\text{b}}$ is the constant background density term. This term is defined to be

\begin{equation}
    \Sigma_{\text{b}} = \frac{n - N^*}{A},
\end{equation}

\noindent where $n$ is the total number of stars in the field of view and $A$ is the total area, normalizing the background density with respect to the selected region.

We use all matched-filter selected sources within a circle of radius 9\arcmin\ surrounding the satellite's initial centroid, projected onto the tangent plane. 
We constrain the centroid to a central box with side lengths of 12\arcmin\ and restrict the half-light radius to a maximum size of 6\arcmin. Constraints for all model parameters can be found in Table \ref{tab:priors}, and we assume flat priors between the presented bounds in all cases.
The \texttt{emcee} program used 64 walkers, each going through 15,000 iterations and the first 7500 iterations were thrown out to account for burn-in. Through this analysis, we find that \boov\ is located at a Right Ascension (RA) of 14h 15m 38.6s and a Declination (Dec) of +32\textdegree\ 54\arcmin\ 40\arcsec. We present the results of the MCMC analysis as a corner plot in Figure \ref{fig:corner} and we include the final parameter estimates alongside all other measured and derived properties of \boov\ in Table \ref{tab:properties}.

\begin{table}
    \centering
    \caption{Flat priors for each parameter in the MCMC analysis.}
    \begin{tabular}{c c}
        \hline
        Parameter & Prior \\
        \hline\hline
        $x_0$ & $-6$\arcmin $<$ $\Delta x_0$ $<$ $+6$\arcmin \\
        $y_0$ & $-6$\arcmin $<$ $\Delta y_0$ $<$ $+6$\arcmin \\
        \rh & 0 $<$ \rh\ $<$ $6$\arcmin \\
        $\epsilon$ & 0 $<$ $\epsilon$ $<$ 1 \\
        $\theta$ & $-$90\textdegree\ $<$ $\theta$ $<$ $+$90\textdegree \\
        $N^*$ & 0 $<$ $N^*$ $<$ 100 \\
        \hline
    \end{tabular}
    \label{tab:priors}
\end{table}

\begin{table*}
    \centering
    \caption{Measured and Derived Properties for \boov}
    \begin{tabular}{c c c}
        \hline
        Property & Description & Value \\
        \hline\hline
        $\alpha_{J2000}$ & Right Ascension & 14h 15m 38.6\,$^{+0.5}_{-0.6}$s \\
        $\delta_{J2000}$ & Declination & +32\textdegree\ 54\arcmin\ 40\,$^{+9.2}_{-10}$\arcsec\\\
        $r_{\text{h, ang}}$ & Angular half-light radius & 0.93$^{+0.26}_{-0.19}$\arcmin \\
        $r_{\text{h, phys}}$ & Physical half-light radius & 26.9$^{+7.5}_{-5.4}$\,pc \\
        $\epsilon$ & Ellipticity & 0.35$^{+0.17}_{-0.21}$ \\
        $\theta$ & Position angle & 30$^{+24}_{-25}$\textdegree \\
        $N^*$ & Number of stars & 37$^{+8.1}_{-7.2}$ \\
        $d$ & Heliocentric distance & 100 $\pm$ 20 kpc \\
        \ebv & Average \ebv, 2\arcmin\,$\times$\,2\arcmin\ box & 0.013 \\
        $M_V$ & Absolute $V$-band magnitude & $-$4.5 $\pm$ 0.4\,mag \\
        $M_r$ & Absolute $r$-band magnitude & $-$4.9 $\pm$ 0.4\,mag \\
        $M_i$ & Absolute $i$-band magnitude & $-$5.2 $\pm$ 0.4\,mag \\
        $\mu_{eff}$ & effective surface brightness & 25.7 $\pm$ 0.7\,mag\,arcsec$^{-2}$ \\
        $\mu_{\alpha}$ cos $\delta$ & Proper Motion in R.A. & $-$0.23 $\pm$ 0.04 (stat) $\pm$ 0.033 (sys) mas\,year$^{-1}$\\
        $\mu_{\delta}$ & Proper Motion in Dec. & $-$0.28 $\pm$ 0.07 (stat) $\pm$ 0.033 (sys) mas\,year$^{-1}$\\
        \hline
        $v_r$ & Radial Velocity [GMOS-N] & +5.1 $\pm$ 13.4\,km\,s$^{-1}$ \\
        $[$Fe/H] & Metallicity [GMOS-N] & $-$2.85 $\pm$ 0.10\,dex\\
        \hline
    \end{tabular}
    \tablecomments{Statistical errors on the proper motion measurements come from the maximum likelihood analysis \\described in Section \ref{subsec:gaia}. Systemic errors were investigated by \citet{Lindegren2021}}
    \label{tab:properties}
\end{table*}

\begin{table*}
    \centering
    \caption{Identifier, Magnitudes, and Distance information for three potential BHB stars}
    \begin{tabular}{c c c c c}
        \hline
        Identifier & $g_0$ (mag) & $r_0$ (mag) & $M_g$ (mag) & $d_{BHB}$ (kpc) \\
        \hline
        \hline
        SDSS J141543.35+325624.0 & 20.42 $\pm$ 0.021 & 20.56 $\pm$ 0.031 & 0.48 $\pm$ 0.058 & 97.4 $\pm$ 2.8 \\
        SDSS J141537.78+325445.0 & 20.43 $\pm$ 0.022 & 20.57 $\pm$ 0.032 & 0.48 $\pm$ 0.069 & 97.4 $\pm$ 3.2 \\
        SDSS J141538.35+325501.0 & 20.72 $\pm$ 0.026 & 21.05 $\pm$ 0.046 & -- & -- \\
        \hline
    \end{tabular}
    \label{tab:BHBs}
\end{table*}

\subsection{Distance Determination}
\label{subsec:dist}

\boov\ was initially identified as an overdensity of stars at around 100\,kpc. Given this initial distance estimate, comparisons to isochrones suggested that there may be a potential member star near the tip of the red giant branch (RGB), as well as three possible member stars consistent with being blue horizontal branch (BHB) stars. These are shown as a red triangle and blue squares respectively, in Figure \ref{fig:field}. Note that the two brightest BHB candidates occupy an almost identical position in this figure.

Under the assumption that the bright star is indeed an RGB star, we can estimate the distance to \boov\ by assuming that this star is at the tip of the RGB (TRGB). This is the upper bound on the distance estimate as the lone, bright star may not truly be at the very tip, and there are no other similarly bright stars in the system with which to compare. 

The TRGB in the $I$-band has been shown to act as a standard candle and can therefore be used for distance estimation \citep{Lee1993, Salaris1997}. As we are working in the PanSTARRS-i band, we estimate the $i$-band TRGB absolute magnitude empirically using PARSEC isochrones \citep{Bressan2012}. Using isochrones of 12 and 13\,Gyr, and considering [Fe/H] in the range [-2.9, -0.9]\,dex in intervals of 0.2\,dex, we find the mean $i$-band TRGB to be $3.52 \pm 0.04$\,mag. We also take into account the statistical uncertainty (0.05\,mag) and the systematic uncertainty (0.08\,mag) from the globular cluster calibration of the $I$-band TRGB by \citet{Freedman2020}, giving $M_i^{\text{TRGB}} = 3.52 \pm 0.10$\,mag.
The corrected $i$-band luminosity of the RGB star in question is $i_0 = 17.05 \pm 0.02$\,mag, yielding a distance modulus of $(m - M)_i = 20.57 \pm 0.10$\,mag. 
The upper bound on the distance to \boov\ is therefore $d_{\text{upper}} = 130.2 \pm 6.1$\,kpc.

For determining the distance to the potential BHB stars, we follow the calibration given by \citet{Deason2011} for the absolute $g$-band magnitude of BHB stars using SDSS $g$- and $r$-band magnitudes:

\begin{eqnarray}
    M_g = 0.434 - 0.169(g_0 - r_0) + 2.319(g_0 - r_0)^2 + \\
    \nonumber 20.449(g_0 - r_0)^3 + 94.517(g_0 - r_0)^4.
\end{eqnarray}

All three potential BHB stars are in SDSS DR17 and their extinction-corrected SDSS $g$ and $r$ magnitudes are presented in Table~\ref{tab:BHBs}. The faintest BHB star has $(g_0 - r_0)$ = $-$0.34, which falls outside the color range for which the above equation is valid ($-0.25$ $<$ $(g_0 - r_0)$ $<$ 0, see \citet{Deason2011} for details), so we cannot use it for a distance estimate. Table~\ref{tab:BHBs} presents derived absolute $g$ magnitudes and distances for the two bright BHBs. They have heliocentric distances of 97.4 $\pm$ 2.8 and 97.4 $\pm$ 3.2\,kpc, where the uncertainties are only those obtained by propagating the uncertainties in magnitudes.

A third estimate of the distance may be measured by estimating the distance an isochrone should be shifted to in order to best match the CMD. As such, we determined the best-match isochrone through visual inspection by adopting a 13\,Gyr stellar population, and considering [Fe/H] in the range [$-2.9$, $-1.9$]\,dex in intervals of 0.2 dex, and distance shifts in the range [80, 175]\,kpc in intervals of 5\,kpc. We found that an [Fe/H] $= -2.3$\,dex isochrone with a distance of 100 $\pm$ 20 kpc describes the data reasonably well. Figure \ref{fig:field} shows the CMD of all stars within 3 elliptical \rh, with the isochrone overlayed at distances of 80, 100, and 120\,kpc, to demonstrate the range we estimate as reasonable: much closer or much more distant, and the isochrone (both the hydrogen and helium burning branches) does not align well with the bulk of the stars, especially, but not only, the BHB and RGB candidates discussed earlier.

While each of these methods of estimating the distance to this satellite are not precise, they all point to a distance of approximately 100\,kpc. Consequently, for the remainder of our analysis, we adopt a distance to the satellite of 100 $\pm$ 20\,kpc. This estimate from the by-eye isochrone analysis is fully consistent with estimates from both BHB stars and the upper limit provided by the candidate RGB star.

We searched the PS1 RR Lyrae \citep{Sesar2017} and Gaia variability \citep{Gaia2022} catalogs, but could not find evidence that any RR Lyrae stars have been detected, so we unfortunately cannot use the properties of these variable stars to constrain the distance to \boov.

\begin{figure*}
    \centering
    \includegraphics[width=\linewidth]{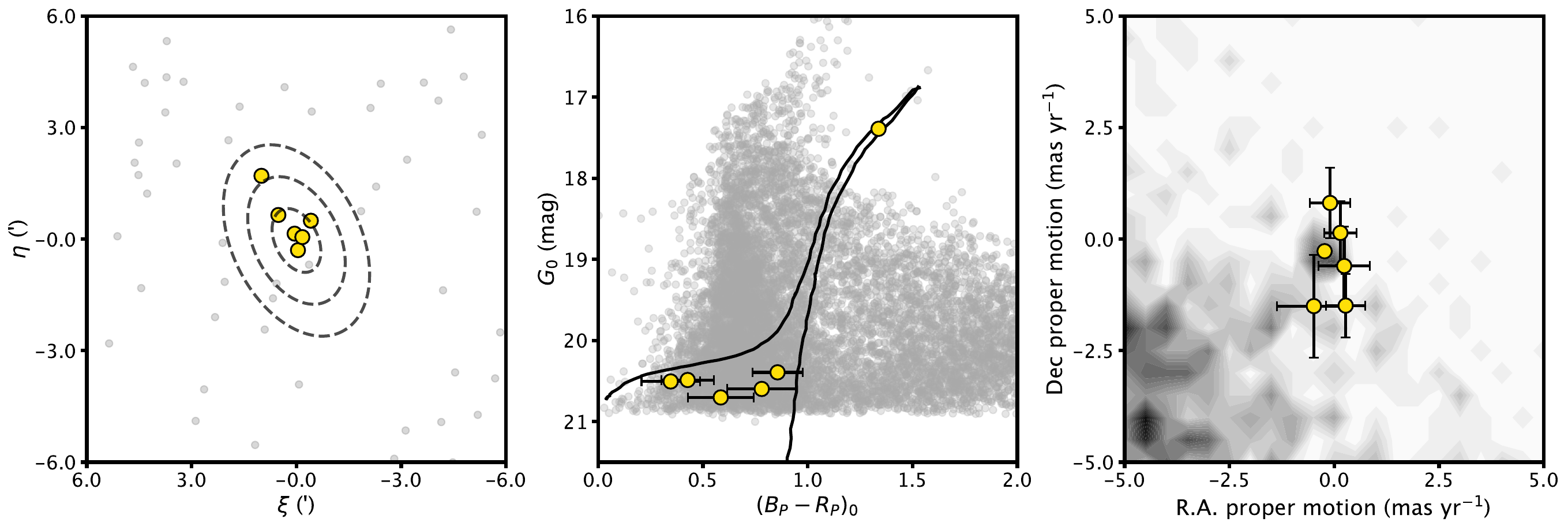}
    \caption{
    Yellow sources are those selected as high confidence (probability $>$ 90\%) members by our maximum-likelihood membership selection algorithm. Grey sources/shading are all 5-parameter Gaia detections in a 2\textdegree\ circle around \boov.
    \plot{Left} Sky positions of stars in a 12\arcmin\,$\times$\,12\arcmin\ region about \boov, projected onto the tangent plane centered at RA, Dec of (14h 15m 38.6s, +32\textdegree\ 54\arcmin\ 42\arcsec). 
    \plot{Center} Color-magnitude diagram with a 13\,Gyr, [Fe/H] = $-$2.3 isochrone overlaid.
    \plot{Right} Proper motions of members overlaid on Milky Way foreground stars.
    }
    \label{fig:gaia}
\end{figure*}

\subsection{Luminosity}
\label{subsec:luminosity}

To estimate the luminosity of \boov, we use an approach that is based on the work of \citet{Martin2016}, which aims to create a suite of synthetic stellar populations that represent the full stellar content of the dwarf galaxy. The construction of each individual synthetic stellar population is performed as follows.

We assume that the stellar population in question is well represented by a 13\,Gyr, [Fe/H] $= -2.3$ PARSEC isochrone \citep{Bressan2012}, with a standard Kroupa initial mass function \citep{Kroupa2001}, that has been corrected for Galactic foreground extinction (same as description in Section \ref{subsec:unions}). We then draw a random distance from a normal distribution centered at 100\,kpc, with a standard deviation of 20\,kpc, and shift the isochrone to that distance. We then construct an $i$-band luminosity function (ignoring the horizontal branch) down to 0.1\,\Msun, and normalize it so that it behaves as a probability distribution function (PDF).

Crucially, these mock stellar populations are characterized by the number of stars above the magnitude limit of our survey, emulating the actual stellar population that was observed. We reran the MCMC analysis using only stars consistent with the main sequence (MS) and red giant branch (RGB) that have Pan-STARRS-i $<$ 24, and we found that \boov\ is estimated to have 34$^{+7.3}_{-6.5}$ stars that fit these criteria. So, we also sample a normal distribution centered at 34, with a standard deviation of 6.9 (average of empirical 16th and 84th quantiles), and set this value to be $N$, the target number of stars in the mock stellar population brighter than the limiting magnitude of 24 mag in Pan-STARRS-i. Randomly drawing values for both the distance and the number of stars above the magnitude limits for each realisation of the stellar population allows for the propagation of parameter estimate uncertainties into the final derived magnitude of the system.

Finally, to create the synthetic stellar population, we sample the PDF using the acceptance-rejection method, recording the $r$-band, $i$-band, and $V$-band magnitudes of each accepted star, and flaging each star that has $m_i < 24$\,mag. When we accrue $N$ flagged stars (where $N$ is the target number of stars in that realization of \boov), we stop sampling. Finally, we shift apparent magnitudes to absolute magnitudes (using the distance corresponding to that specific realization of the stellar population), convert to fluxes, sum them, and convert to the total absolute magnitude.

We generate 1000 instances of the stellar population and find the systemic magnitude of \boov\ to be $M_V = -4.5$ $\pm$ $0.4$\,mag. When converted into total luminosity, we get $5.4^{+2.2}_{-1.6}$\,$\times$\,$10^3$\,\Lsun. We calculate the effective surface brightness by dividing half the total flux by the area enclosed by one elliptical half-light radius and converting to mag\,arcsec$^{-2}$. We calculate it to be 25.7 $\pm$ 0.7\,mag\,arcsec$^{-2}$ with all errors from the magnitude, half-light radius, and ellipticity propagated through. These values, including the absolute magnitudes in the $r$- and $i$-bands, are included in Table~\ref{tab:properties}.

\section{Dynamics and Metallicity}
\label{sec:dynmet}

\subsection{Proper Motion, Membership using Gaia}
\label{subsec:gaia}

The Third Data Release from Gaia \citep{Gaia2016, Gaia2021} has a limiting magnitude of about $G = 21$\,mag. 
The majority of stars we selected as members of this system based on UNIONS photometry using our matched-filter method are fainter than the Gaia limits, but several of the brighter stars are successfully detected by Gaia, allowing for the use of Gaia's powerful astrometry \citep{Lindegren2021} in characterizing this system, as well as providing clear confirmation of its reality.

We follow the methodology developed by \citet{McConnachie2020a} to estimate the systemic proper motion of the system and to clearly identify member stars. The reader is referred to this paper for details. Briefly, the algorithm estimates the most likely proper motion for a putative satellite under the assumption that a field of stars consists of only a Milky Way satellite and Milky Way foreground, by considering the CMD distribution of stars, the spatial distribution of stars, and the proper motion distribution of stars. The foreground/background density is assumed constant in the region of the putative satellite, and the CMD and proper motion distributions of the foreground/background are derived empirically. The CMD distribution of the putative satellite assumes an old, metal-poor system at some distance, and the spatial distribution of the putative satellite assumes a 2D exponential profile described by a centroid, ellipticity, position angle and half-light radius. As such, the distance and structural parameter estimates derived earlier (and their uncertainties) are used in this analysis. The CMD distribution is modelled as an old (13\,Gyr), metal poor ([Fe/H] = $-$2.3) isochrone in the Gaia passbands \citep{Riello2021}. Additionally, we incorporated extra constraints for robustly selecting horizontal branch stars (see Jensen et al. 2022, in prep, for further details).
This method assumes that the proper motions of all stars within the putative satellite share the same intrinsic proper motion, with any variance coming directly from measurement uncertainties, so that the proper motion PDF is modeled as a bivariate Gaussian function.

For \boov, we follow \citet{McConnachie2020b}, and select well-measured sources in a circular region with a radius of 2\textdegree\ around the previously determined centroid. We identify six high confidence members (membership probability $>$ 90\%) within 3 elliptical \rh\ and display these stars in Figure \ref{fig:gaia}. This figure shows the spatial positions, $G_0$ versus $(B_P-R_P)_0$ CMD, and proper motions of the 6 members compared to the field (grey points). We find the systemic proper motion of \boov\ to be $(\mu_\alpha$cos$\delta, \mu_\delta) = (-0.23$ $\pm$ 0.04 (stat) + 0.033 (sys), $-0.28$ $\pm$ 0.07 (stat) + 0.033 (sys)) mas year$^{-1}$. We note that such a tight clustering in all these spaces, especially proper motion space, provides a robust confirmation that this is a real physical system. Systematic errors are taken from \citet{Lindegren2021}.

Encouragingly, all of these high confidence members are also found in the original matched-filter selection that identified the stellar overdensity in UNIONS (Section \ref{subsec:detection}). Of these six likely member stars, two are the brightest stars identified as potential BHB stars in Section \ref{subsec:dist}. The third BHB star, which was not used for a distance estimate, does not have well measured astrometric parameters, and so its membership cannot be classified by this method. Another of the six likely member stars is the bright RGB candidate discussed earlier. The Gaia member stars are cross-matched to the UNIONS data set and are plotted as yellow markers on Figure \ref{fig:cmdselect}. We note that in Figure \ref{fig:gaia}, several of the faintest stars appear to be HB stars that lie below the isochrone track. While this might suggest we are underestimating the distance, shifting the isochrone to larger distances only produces a reasonable fit to the data up to about 120\,kpc, which is within the uncertainty we adopt when only using UNIONS data. A deeper CMD than obtained with Gaia or UNIONS is necessary to allow  a more robust distance measurement.

\subsection{Spectroscopic Follow-up Analysis}
\label{subsec:spec}

\begin{figure*}
    \centering
    \includegraphics[width=0.8\linewidth]{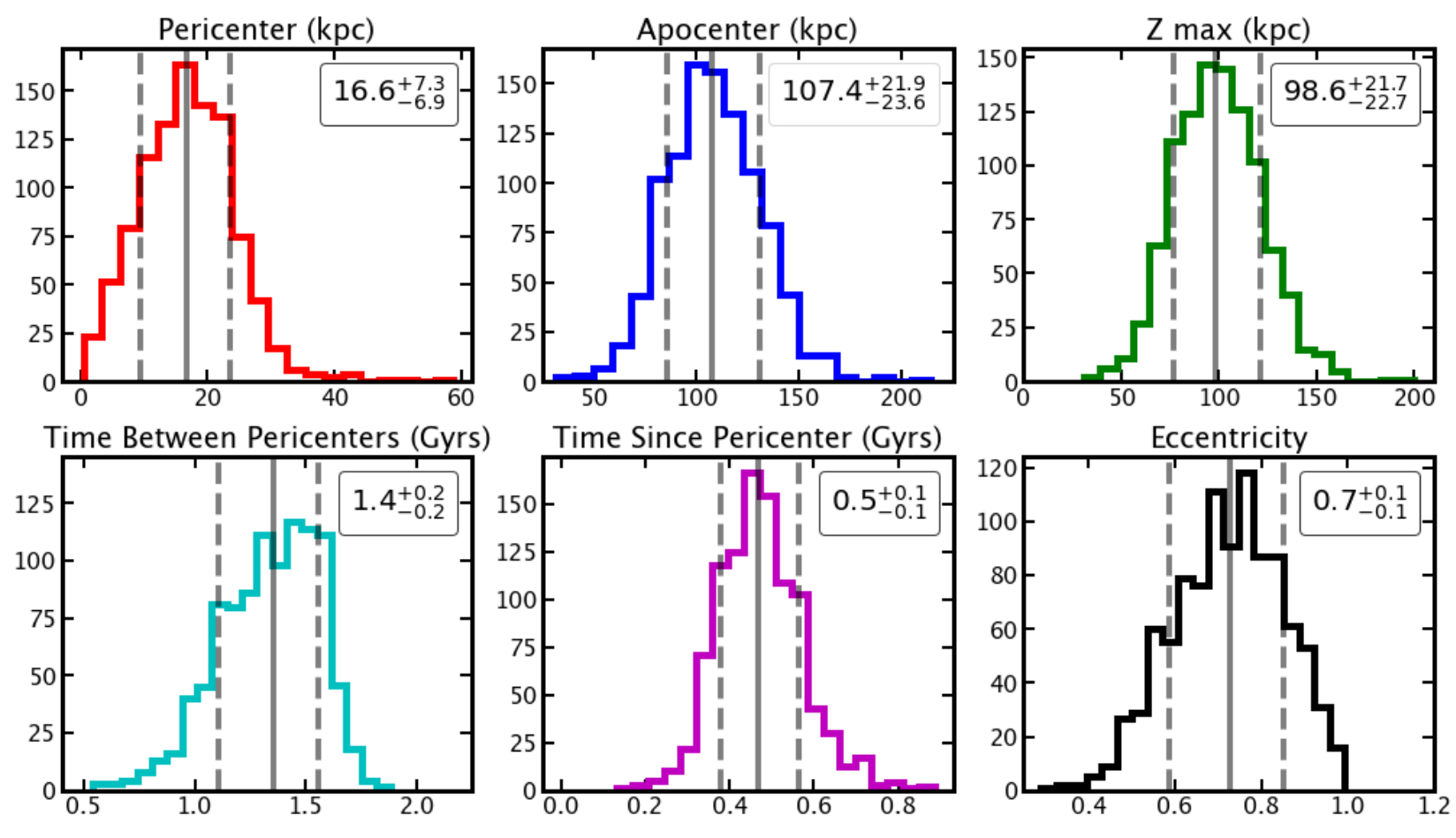}
    \caption{Resulting 1D distributions of key parameters from orbital MCMC analysis. It is estimated to take $\sim$1.4\,Gyr for \boov\ to travel between pericenters and it has been $\sim$0.5\,Gyr since last pericenter.
    }
    \label{fig:orbitdist}
\end{figure*}

In Section \ref{subsec:dist} we assumed that the brightest star consistent with the isochrone was a red giant star near the TRGB. Our membership analysis has also identified this star as a highly likely member (probability $> 99.9\%$). Fortuitously, this star was observed as part of the Large Sky Area Multi-Object Fibre Spectroscopic Telescope (LAMOST), and has low-resolution spectroscopy available as part of its Data Release 5 (DR5). Its LAMOST stellar identifier is LAMOST HD141746N331518M01. LAMOST DR5 \citep{Xiang2019} provides an estimation for both the metallicity and heliocentric radial velocity of this RGB star, with [Fe/H] = $-$2.25 $\pm$ 0.60\,dex $v_r$ = $-5.5 \pm 22.7$\,\kms. However, the uncertainties on both measurements are rather large, which prompted us to obtain our own low-resolution spectrum of this RGB star using GMOS-N as described in Section \ref{subsec:gmos}. The analysis of the reduced spectrum is now presented.

We infer the metallicity of the star adopting the method from \citet{Starkenburg2010}. Their method needs as input the equivalent width (EW) of the second and third components of the Ca\ii{} Triplet ($\lambda\lambda 8498.02, 8542.09, 8 662.14$ \AA) and the absolute magnitude of the star $M_V$ \citep[see Eq.A.1 from][]{Starkenburg2010}. The EW is measured using the \textsc{splot} routine in \textsc{IRAF} \citep{Tody86,Tody93} fitting with multiple line profiles. The median and the standard deviation has been adopted as final values for the EW and its uncertainty. $M_V$ is derived by converting the Gaia DR3 magnitudes to the Johnson-Cousin filter following the relation from \citet{Riello2021} and adopting a heliocentric distance of $100$ $\pm$ $20$\,kpc. We perform a Monte Carlo analysis with $10^6$ randomizations on the heliocentric distance, the EW, and the de-reddened magnitudes assuming Gaussian distributions for all parameters. The final [Fe/H] and its uncertainty are the median and the standard deviation of the randomizations, respectively. The star is measured to be a very metal-poor star, with [Fe/H] = $-$2.85 $\pm$ 0.10\,dex.

To calculate the radial velocity, we create a synthetic spectrum using the \textit{synth} option in \textsc{MOOG} \footnote{\url{https://www.as.utexas.edu/~chris/moog.html}} \citep{Sneden73} with the list of spectral lines generated by \textsc{linemake}\footnote{\url{https://github.com/vmplacco/linemake}} \citep{Placco21}. We adopted a model atmosphere from the MARCS1 models \citep{Gustafsson08,Plez12}. The synthetic spectrum has been created at the same resolution of the GMOS spectrograph with the stellar parameters $\rm{T_{eff}}$, logg, and [Fe/H], where each parameter is derived with the following methods. The effective temperature is derived using the calibration from \citet{Mucciarelli21}, which takes as input the knowledge on the nature of the star (\ie dwarf or giant), the metallicity, and the de-reddened Gaia EDR3 photometry. Then, the surface gravity is inferred adopting the Stefan-Boltzmann equation. This needs the distance, the de-reddened G magnitude, the bolometric corrections on the flux \citep{Andrae18}, an estimate on the effective temperature, and the stellar mass ($0.5-0.8$\,\Msun). A Monte Carlo randomization has been performed to infer the $\rm{T_{eff}}$, logg, and their uncertainties. These methods have proved to yield robust stellar parameters compatible with high-resolution spectroscopic values in the very metal-poor regime \citep[\eg][]{Kielty21,Sestito22,Waller22}. We derive $\rm{T_{eff}}= 4481$ $\pm$ $77$\,K and logg $=0.87$ $\pm$ $0.11$, and incorporate these parameters into the synthetic spectrum. Then, we added a Poissonian noise to match the observations. The radial velocity is measured by cross-correlating the combined observed spectrum with the downgraded synthetic spectrum using the \textsc{fxcor} routine in \textsc{IRAF}, and is found to be 5.1 $\pm$ 13.4\,\kms. In the absence of any other stars in \boov\ with measured radial velocities, we adopt this radial velocity as representative of the systemic velocity of the candidate dwarf.

\subsection{Orbital analysis}
\label{subsec:orbit}

We now use a simple dynamical model to examine the orbit of \boov, with the aim of understanding its orbital history and interaction with the Milky Way. We approximate \boov\ as a point-mass in a Milky Way potential, implemented with the \textsc{python}-wrapped package \textsc{gala} \citep{pricewhelan2017}. The Milky Way potential used for this analysis is identical to that which is described in Section~5 of \citet{Jensen2021}, so we direct the reader to that work, and references therein, for full details.

We perform 1000 realizations of the point-mass orbit, where the initial conditions (i.e. input parameters \{~$\alpha_{J2000}$, $\delta_{J2000}$, $d$, $\mu_{\alpha}$ cos$\delta$, $\mu_{\delta}$, $v_r$~\}) for each realization are drawn from normal distributions with means and standard deviations defined by the values and associated errorbars presented in Table \ref{tab:properties}.

Each point-mass is integrated both forwards and backwards in time by 1\,Gyr in time steps of 10$^{-3}$\,Gyr. For each orbit, the pericenter (closest approach to Milky Way), apocenter (furthest point from Milky Way), Z max (maximum height above the disk), time between pericenters (orbital time), time since last pericenter, and orbital eccentricity are recorded and presented in 1D distributions in Figure \ref{fig:orbitdist}.

\begin{figure}
    \centering
    \includegraphics[width=\linewidth]{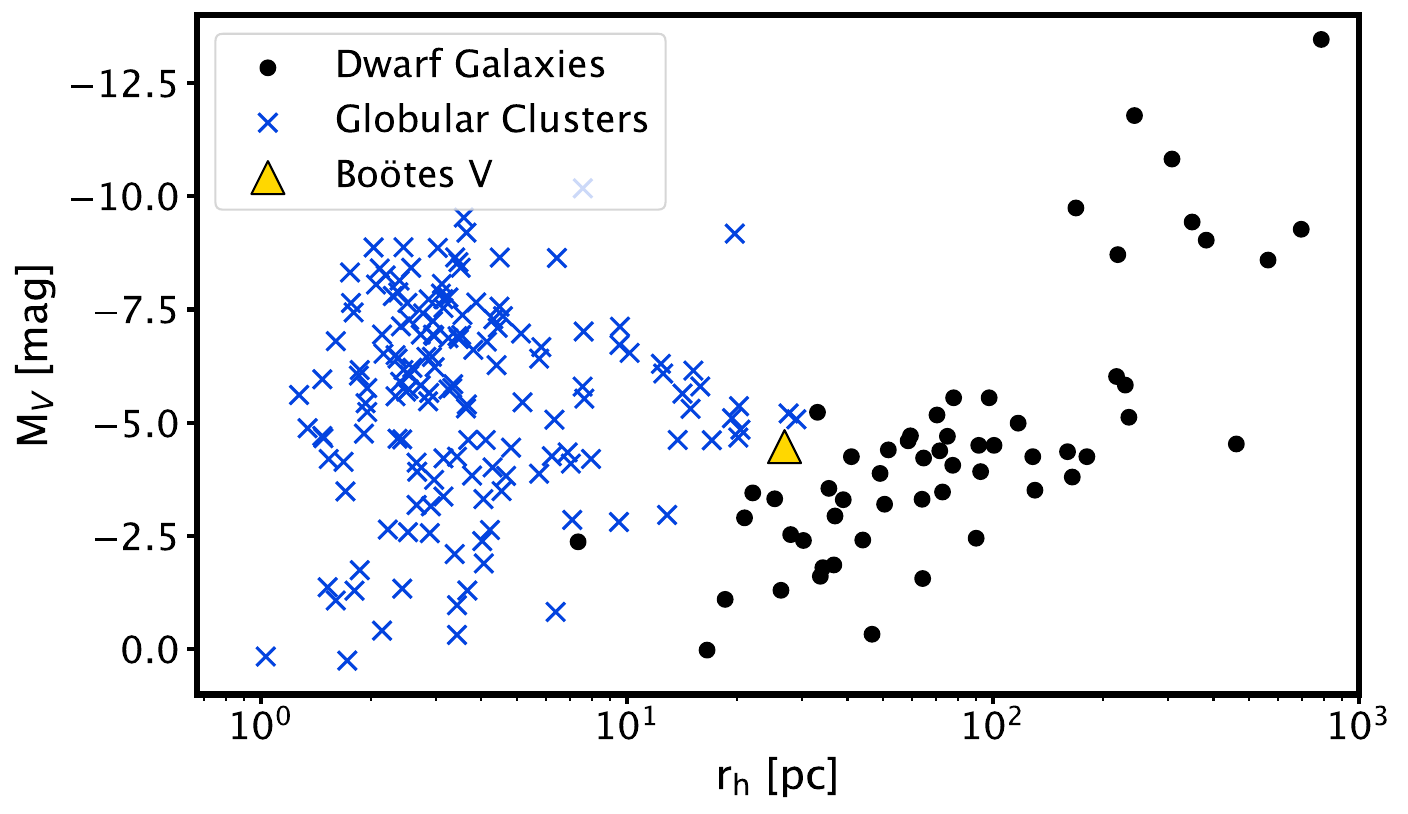}
    \caption{Absolute V-band magnitude ($M_V$) vs. half-light radius (\rh) plane showing both the dwarf galaxy (black markers) and globular cluster (blue Xs) populations. We show both candidate and spectroscopically confirmed dwarf galaxies within 300\,kpc of the Milky Way. \boov\ is shown as a yellow triangle.}
    \label{fig:rhMv}
\end{figure}

\section{Classification and Conclusions}
\label{sec:conc}

In this paper we have presented the discovery of the \boov\ satellite of the Milky Way in UNIONS. We detected it using a matched-filter method on the combined CFIS-r and Pan-STARRS-i photometric catalog, from which we estimated the distance, luminosity and structural parameters of the system. Examination of the Gaia DR3 data allows us to confirm this as a real system, identify its brightest members in the Gaia data, and estimate its proper motion. GMOS-N observations also allow us to provide a first estimate of its spectroscopic metallicity and its radial velocity, which in turn allow us to characterize the orbit of the system.

Based on the results of the orbital estimation, \boov\ is approaching the apocenter of its orbit, having been at pericenter $\sim$0.5\,Gyr ago. Interestingly, the pericenter of the orbit is 16.6$^{+6.8}_{-7.6}$\,kpc from the center of the Milky Way. With an orbital period of $1.4$\,Gyr, \boov\ may have had time to complete several orbits. Several such close passes with the inner part of the Milky Way potential may mean that tidal forces have affected this satellite, though there does not appear to be significant visual evidence in support of tidal disruption. Additionally, the orbit is highly eccentric and the motion of the satellite is primarily occurring perpendicular to the disk.

We note, however, that the measure of a single radial velocity is a poor proxy for the systemic radial velocity of the satellite, as both dwarf galaxies and globular clusters are measured to have velocity dispersions on the scale of a few to tens of \kms. We also note that the observed star for which we obtained spectroscopy is a visual double, with a companion that does not have fully measured astrometric parameters in Gaia DR3. However, the double is separated on the sky by 1.72\,\arcsec, which, at a distance of 100\,kpc, corresponds to a minimum physical separation of 0.83\,pc. It is very unlikely that these two stars are bound, and if they are, a negligible portion of the radial velocity on the measured star would be due to binary orbital motion. Regardless, the radial velocity of this single, bright RGB star is still our best available tool for providing an initial estimate of the orbit for \boov.

\boov\ is faint ($M_V$ = $-$4.6 $\pm$ 0.4\,mag), small (\rh\ = 27.1$^{+7.2}_{-5.6}$\,pc), and distant ($d$ = 100 $\pm$ 20\,kpc). A question remains as to whether this system is a dwarf galaxy or a globular cluster. Figure \ref{fig:rhMv} illustrates the absolute V-band magnitude ($M_V$) versus half-light radius (\rh) plane, a commonly used metric for sorting dwarf galaxies and globular clusters. Here, we use only dwarf galaxies within 300\,kpc of the Milky Way. The data and separation are taken from the updated dataset of \cite{McConnachie2012}\footnote{\url{https://www.cadc-ccda.hia-iha.nrc-cnrc.gc.ca/en/community/nearby/}}. For the globular clusters, we used data from \cite{Baumgardt2021}\footnote{\url{https://people.smp.uq.edu.au/HolgerBaumgardt/globular/parameter.html}}. \boov\ resides in a region of this plane that has largely contained satellites whose classification has been contested. However, it is worth noting that, in terms of scale-radius (\rh), \boov\ would be the third largest globular cluster while only being the seventh smallest dwarf. Additionally, at a heliocentric distance of 100\,kpc, \boov\ would also be the fourth most distant globular cluster, but at a distance typical of dwarfs.

The metallicity of this system is very notable. The GMOS-N spectrum was measured to be very metal-poor, with an [Fe/H] = $-$2.85\,dex, indicating that \boov\ may be among the most metal-poor dwarf galaxies known. In terms of overall mean metallicities,  Tucana II is the only dwarf with a mean metallicity more metal poor than this, at $-$2.90\,dex \citep{Chiti2018}. Four additional dwarf galaxies have $\langle$[Fe/H]$\rangle$ $< -2.7$ \citep[Bo\"otes II, Draco II, Horologium I, \& Segue 1;][although the classification of Draco II is unclear at this time]{Ji2016, Longeard2018, Kosposov2015b, Frebel2014}. Each of these five very metal-poor dwarfs contain individual stars that are measured to have [Fe/H] $\lesssim$ -3\,dex. While this single metallicity measurement certainly suggests \boov\ must be quite metal poor, we do not yet know how representative this star is of the overall mean metallicity of the system, and more spectroscopic observations are needed.

Globular clusters do not appear to extend to such low mean metallicities as dwarf galaxies. Indeed, if \boov\ were a globular cluster, we would expect this single measurement to be more representative of the whole system, and it would be nearly 0.5\,dex more metal-poor than then most extreme, {\it intact} globular cluster \citep[][2010 edition]{Harris1996}. However, it is worth noting that two remarkably metal-poor stellar streams have been discovered whose progenitors are thought to be ancient globular clusters, the Phoenix stellar stream \citep[][]{Wan2020}, and the C-19 stellar stream \citep[][]{Martin2022}, which have [Fe/H] $\simeq -$2.70\,dex and [Fe/H] $\simeq -$3.38\,dex, respectively.

The rather large caveat throughout this entire discussion is that we only have a metallicity measurement for a single star. However, it does immediately make \boov\ an interesting object of future spectroscopic studies as results from the observations of several more stars will be two-fold: measurements of its stellar velocity dispersion will provide a direct estimation of the mass (and by extension, the dark matter content), and measurements of the metallicity dispersion have more recently been used as a proxy for mass, insofar as systems with a large metallicity dispersion likely need to reside, or have resided, in a massive dark matter halo \citep[e.g.][]{Willman2012}. Both measurements will likely provide clarity relating to the classification of this newly discovered system. While it is not inconceivable that this system could be a globular cluster, we consider the balance of evidence to currently suggest that \boov\ is more likely a member of the ever-growing class of ultra-faint dwarf galaxies.

\boov\ is the first Milky Way satellite to be discovered in the UNIONS dataset, which, when complete, will provide coverage of approximately 5000 square degrees of the northern extragalactic sky at a depth comparable to the first year of the Legacy Survey of Space and Time by the Vera C. Rubin observatory.
The discovery of a single new candidate dwarf galaxy is, in some sense, another drop in the bucket, as the list of known Local Group galaxies continues to grow quickly. 
Even though each new object is worthy of study in its own right, there is a more comprehensive goal underlying all of this work.
The true power of broad searches for new dwarf galaxies is that they build towards amassing a statistically significant population of faint stellar systems whose chemical, dynamical, and structural properties will test theories of dark matter and of galaxy formation on the smallest scales.

\section{Ackownledgements}


We acknowledge and respect the l\textschwa\textvbaraccent {k}$^{\rm w}$\textschwa\ng{}\textschwa n peoples on whose traditional territory the University of Victoria stands and the Songhees, Esquimalt and $\ubar{\rm W}$S\'ANE\'C  peoples whose historical relationships with the land continue to this day.

It was a pleasure to coordinate the submission of this paper with an independent discovery paper led by William Cerny and the DELVE team, and we thank William, Alex Drlica-Wagner, and the whole DELVE team for the very positive, collaborative interactions that we had with them.

We thank the Reviewer whose comments and feedback helped improve the clarity and readability of this manuscript.

This work is based on data obtained as part of the Canada-France Imaging Survey, a CFHT large program of the National Research Council of Canada and the French Centre National de la Recherche Scientifique. Based on observations obtained with MegaPrime/MegaCam, a joint project of CFHT and CEA Saclay, at the Canada-France-Hawaii Telescope (CFHT) which is operated by the National Research Council (NRC) of Canada, the Institut National des Science de l’Univers (INSU) of the Centre National de la Recherche Scientifique (CNRS) of France, and the University of Hawaii. This research used the facilities of the Canadian Astronomy Data Centre operated by the National Research Council of Canada with the support of the Canadian Space Agency. This research is based in part on data collected at Subaru Telescope, which is operated by the National Astronomical Observatory of Japan. We are honored and grateful for the opportunity of observing the Universe from Maunakea, which has the cultural, historical and natural significance in Hawaii. Pan-STARRS is a project of the Institute for Astronomy of the University of Hawaii, and is supported by the NASA SSO Near Earth Observation Program under grants 80NSSC18K0971, NNX14AM74G, NNX12AR65G, NNX13AQ47G, NNX08AR22G, YORPD20\_2-0014 and by the State of Hawaii. 

We thank the staff of Gemini Observatory North for their support in the acquisition of some of these data. We are also grateful to the director of Gemini for granting us Directors Discretionary time to complete this study. These data are based on observations obtained at the international Gemini Observatory, a program of NSF’s NOIRLab, which is managed by the Association of Universities for Research in Astronomy (AURA) under a cooperative agreement with the National Science Foundation on behalf of the Gemini Observatory partnership: the National Science Foundation (United States), National Research Council (Canada), Agencia Nacional de Investigaci\'{o}n y Desarrollo (Chile), Ministerio de Ciencia, Tecnolog\'{i}a e Innovaci\'{o}n (Argentina), Minist\'{e}rio da Ci\^{e}ncia, Tecnologia, Inova\c{c}\~{o}es e Comunica\c{c}\~{o}es (Brazil), and Korea Astronomy and Space Science Institute (Republic of Korea).

This work has made use of data from the European Space Agency (ESA) mission
{\it Gaia} (\url{https://www.cosmos.esa.int/gaia}), processed by the {\it Gaia}
Data Processing and Analysis Consortium (DPAC,
\url{https://www.cosmos.esa.int/web/gaia/dpac/consortium}). Funding for the DPAC
has been provided by national institutions, in particular the institutions
participating in the {\it Gaia} Multilateral Agreement. 

Guoshoujing Telescope (the Large Sky Area Multi-Object Fiber Spectroscopic Telescope LAMOST) is a National Major Scientific Project built by the Chinese Academy of Sciences. Funding for the project has been provided by the National Development and Reform Commission. LAMOST is operated and managed by the National Astronomical Observatories, Chinese Academy of Sciences. 

AWM acknowledges support from the NSERC Discovery Grant program. NFM gratefully acknowledges support from the French National Research Agency (ANR) funded project ``Pristine'' (ANR-18-CE31-0017) along with funding from the European Research Council (ERC) under the European Unions Horizon 2020 research and innovation programme (grant agreement No. 834148). FS thanks the Dr. Margaret "Marmie" Perkins Hess postdoctoral fellowship for funding his work at the University of Victoria.

%

\vspace{5mm}
\facilities{CFHT, PS1, Gemini:Gillett, {\it Gaia}}


\software{\texttt{astropy} \citep{Astropy2013, Astropy2018, Astropy2022}, \texttt{emcee} \citep{ForemanMackey2013}, \texttt{numpy} \citep{numpy2020}, \texttt{scipy} \citep{SciPy2020}}




\bibliography{ref}{}
\bibliographystyle{aasjournal}



\end{document}